\documentclass[12pt]{article}
\usepackage{amsmath}
\usepackage{amssymb}
\usepackage{epsfig}
\usepackage{epic}

\numberwithin{equation}{section}
\numberwithin{figure}{section}

\begin{document}

\def\k{\kappa}
\def\half{\fraction{1}{2}}
\def\fraction#1#2{ { \scriptstyle \frac{#1}{#2} }}
\def\der#1{\frac{\partial}{\partial #1}}
\def\d{\partial}
\def\p2{\fraction{\pi}{2}}
\def\s{\sigma}
\def\t{\tau}
\def\Id{|I\rangle}
\def\Tr{\mathrm{Tr}}
\def\z{\xi}
\def\eq#1{eq.(\ref{eq:#1})}
\def\Eq#1{Eq.(\ref{eq:#1})}

\begin{titlepage}
\rightline{\today}

\begin{center}
\vskip 1.0cm
{\Large \bf{Split String Formalism and the Closed String Vacuum, II} }
\\
\vskip 1.0cm

{\large Theodore Erler}
\vskip 1.0cm
{\it {Harish-Chandra Research Institute} \\ 
{Chhatnag Road, Jhunsi, Allahabad 211019, India}}
\\ E-mail:terler@mri.ernet.in \\

\vskip 1.0cm
{\bf Abstract}
\end{center}

\noindent
In this paper we consider a class of generalizations of Schnabl's solution 
of open bosonic string field theory obtained by replacing the wedge state by
an arbitrary combination of wedge states. We find that under a few modest 
conditions such generalizations give a sensible deformation of Schnabl's 
solution for the closed string vacuum---in particular, we can compute their 
energies and show that they reproduce the value predicted by Sen's conjecture.
Though these solutions are apparently gauge equivalent, they are not in 
general related by midpoint-preserving reparameterizations.

\medskip

\end{titlepage}

\newpage

\baselineskip=18pt

\tableofcontents

\section{Introduction}

Since Schnabl constructed his analytic solution of open bosonic string 
field theory (OSFT)\cite{Schnabl}, it has become possible to imagine that 
OSFT is a background independent framework which could allow exact computations
of the properties of vacua in string theory (see ref.\cite{others} for related
important developments). Given this new solution, 
it is important to explore deformations in the hope of clarifying its 
structure and generating new solutions. As discussed in 
a previous paper\cite{Erler}(henceforth, (I)), the split string 
formalism\cite{Gross-Taylor,Moyal,Okawa} proves immensely useful in this 
capacity. One generalization suggested by this approach 
involves a choice of projector and its associated conformal 
frame\cite{Erler,RZ,RZO}; these new solutions are related to Schnabl's by
midpoint preserving reparameterizations\cite{RZO}.

The split string formalism suggests a second interesting 
generalization\cite{Okawa}: Replace the wedge states separating the 
operator insertions in Schnabl's solution by general sums of wedge 
states. In this paper we consider these solutions and the delicate limiting
process required to calculate their energies. With the properly defined 
limit and some fortuitous cancellations, we are able to demonstrate the 
existence of an infinite class of new exact solutions describing the closed 
string vacuum. These solutions can be partitioned into families labeled by a 
wedge state $\Omega^\gamma$, interpolating between ``identity based'' 
solutions $\gamma=0$ and ``sliver based'' solutions $\gamma=\infty$. The 
solutions are apparently gauge equivalent, but are not in general related by 
reparameterizations.

This paper is organized as follows. In section \ref{sec:setup} we review the 
split string formulation of Schnabl's solution and explain the constraints 
on the wedge state combinations necessary to get nontrivial and well-defined
solutions. In section \ref{sec:energy} we evaluate the energy of these 
solutions and show that they reproduce the correct D-brane tension. In section 
\ref{sec:cohomology} we show that the perturbative spectrum around these 
solutions is empty, along the lines of refs.\cite{cohomology1,cohomology2}. 
We end with some conclusions.

\section{Setup}
\label{sec:setup}

Let us quickly review the construction of Schnabl's solution in the split
string formalism\cite{Okawa,Erler}. In this paper, all of the solutions we 
are interested in take the form,
\begin{equation}\Psi = Fc\frac{KB}{1-F^2}c F\label{eq:SSsol}
\end{equation}
where $F,K,B,c$ are string fields and all products above are 
open string star products. In general $F$ can be any function of $K$,
$$F=F(K)$$
and the fields $K,B,c$ satisfy the simple properties,
\begin{eqnarray}&\ & Bc + cB =1\ \ \ \ \ \ \ \ KB - BK = 0 \ \ \ \ \ \ \ \ 
B^2 = c^2 = 0
\nonumber\\ &\ & \ \ \ 
dK = 0 \ \ \ \ \ \ \ \ \ \ \ \ \ \ \ \ dB = K \ \ \ 
\ \ \ \ \ \ \ \ \ \ \ 
dc = cKc\label{eq:SSid}\end{eqnarray}
where $d=Q_B$ is the BRST operator. A short calculation then shows that $\Psi$
satisfies the equations of motion:
\begin{equation}d\Psi + \Psi^2 = 0\end{equation}
To get Schnabl's original solution, we must make some additional choices in
the definitions of $F,K,B$ and $c$. As explained in (I), an explicit 
realization of the identities \eq{SSid} is equivalent to the choice of a 
projector and its associated conformal frame (see also \cite{RZ,RZO}). 
Schnabl's solution is formulated in the sliver frame, where explicitly 
\begin{eqnarray}
K &=& \frac{\pi}{2}(K_1)_L\Id\ \ \ \ \ K_1 = L_1+L_{-1}\nonumber\\ 
B &=& \frac{\pi}{2}(B_1)_L\Id\ \ \ \ \ B_1 = b_1+b_{-1}\nonumber\\ 
c &=& \frac{1}{\pi}c(1)\Id
\end{eqnarray}
$\Id$ is the identity string field and the subscript $L$ denotes taking
the left half of the corresponding charge (integrating the current 
{\it clockwise} on only the positive half of the semicircle). For 
definiteness, we will focus on solutions in the sliver frame---at any rate, 
the energy calculation in the next section proceeds identically regardless of 
the choice of projector. To get Schnabl's solution we must also make a 
particular choice for $F$:
\begin{equation}F = e^{K/2} = \Omega^{1/2}\end{equation}
where $\Omega^{1/2}$ is the square root of the $SL(2,\mathbb{R})$ 
vacuum\footnote{If we had chosen another projector conformal frame, 
$e^{K/2}$ would not be a wedge state but a member of the abelian algebra of 
interpolating states for some other projector\cite{RZ}.}. 

As discussed in (I) and ref.\cite{RZO}, we may generalize Schnabl's solution
by allowing $F$ to be an arbitrary wedge state,
\begin{equation}F = e^{\gamma K/2} = \Omega^{\gamma/2}\label{eq:magnified}
\end{equation}
for $\gamma\in[0,\infty]$. In the conformal field theory picture, this 
corresponds to ``magnifying'' the strips inside the cylindrical correlators
defining Schnabl's solution.  We will call these {\it pure wedge 
solutions}; they are all related by a simple midpoint preserving 
reparameterization.

In this paper we consider {\it composite wedge solutions}, i.e.
solutions where $F$ is allowed to be a more-or-less arbitrary function of 
$K$. To be specific, we will assume $F$ can be written as a kind of 
``Laplace transform,''
\begin{equation}F = \Omega_f\equiv\int_{-\infty}^{\infty}dt f(t)\Omega^t
\end{equation}
where $f$ is a real function, subject to a few conditions which we will 
explain shortly. Pure wedge solutions are a special case with,
\begin{equation}f(t) = \delta\left(t-\frac{\gamma}{2}\right)\end{equation}
The fields $\Omega_f$ satisfy some nice properties:
\begin{equation}\Omega_f\Omega_g = \Omega_{f\star g}\ \ \ 
\Omega_f+\Omega_g = \Omega_{f+g}\end{equation}
where $f\star g$ is the convolution product:
\begin{equation}f\star g(t) =\int_{-\infty}^\infty ds f(t-s)g(s)
\end{equation} The identity of the convolution is a delta function, 
$1_\star=\delta(t)$.  

\begin{figure}[top]
\begin{center}
\resizebox{3.6in}{1.9in}{\includegraphics{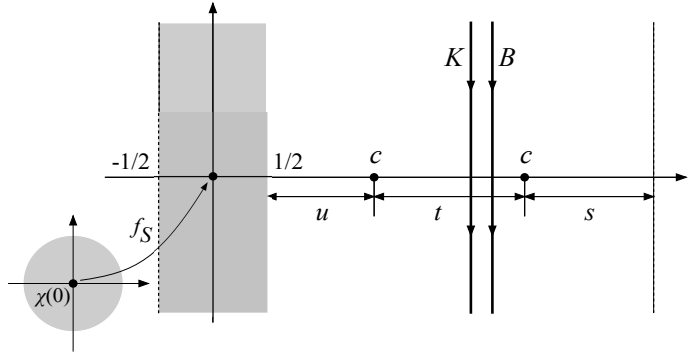}}
\end{center}
\caption{\label{fig:BBlocks}
Conformal field theory representation of the state 
$\Omega^s cKB\Omega^t c\Omega^u$.}
\end{figure}

The solution \eq{SSsol} can be rewritten,
\begin{equation}\Psi = \Omega_f cKB 
\Omega_{\left(\frac{1}{1-f^2}\right)_\star} c\Omega_f\label{eq:O_fsol}
\end{equation}
where all the function products in the subscript are calculated with
convolution. This is basically a shorthand notation for a triple 
integral over states of the form,
\begin{equation}\Omega^scKB \Omega^t c\Omega^u \label{eq:BBlocks}\end{equation}
In conformal field theory language, these states can be defined in terms of
correlation functions on the cylinder $C_{s+t+u+1}$ with particular 
insertions. The cylinder is a strip in the complex plane 
$-\frac{1}{2}<\Re(z)<s+t+u+\frac{1}{2}$ with its boundaries identified; 
the local coordinate occupies the region $-\frac{1}{2}<\Re(z)<\frac{1}{2}$ 
and is the image of the unit disk under the sliver conformal map,
\begin{equation}f_\mathcal{S}(z) = \frac{2}{\pi}\tan^{-1}z\end{equation}
Explicitly, \eq{BBlocks} is defined via the correlator,
\begin{equation}\langle\Omega^scKB \Omega^t c\Omega^u,\chi\rangle
=\left\langle c\left(t+u + \frac{1}{2}\right)KB
c\left(u+\frac{1}{2}\right)f_\mathcal{S}\circ\chi(0)
\right\rangle_{C_{s+t+u+1}}
\end{equation}
where $K,B$ are the contour insertions,
\begin{equation}K = \int_{i\infty}^{-i\infty}\frac{dz}{2\pi i}T(z)\ \ \ \ 
B = \int_{i\infty}^{-i\infty}\frac{dz}{2\pi i}b(z)
\label{eq:KB_cont}\end{equation} 
to be integrated between the two $c$ ghosts (see figure \ref{fig:BBlocks}).

Let us now state the conditions which must be imposed on defining function
$f(t)$:
\begin{eqnarray}\mathbf{(i)} &\ &\ \ f(t) = 0\ \mathrm{for}\ t<0
\nonumber\\
\mathbf{(ii)}&\ &\ \ \int_{-\infty}^\infty dt f(t) = \sqrt{\lambda} = 1
\nonumber\\
\mathbf{(iii)}&\ &\ \ \int_{-\infty}^\infty dt\, tf(t) = \frac{\gamma}{2} >0
\end{eqnarray}
Condition {\bf (i)} follows from general considerations, specifically that
the solution must not involve inverse wedge states. If such states exist, 
they are singular and best avoided. Condition {\bf (ii)}, as we 
will see, implies that the solution produces the energy of a single 
D-brane, in accordance with Sen's conjectures. The constant $\lambda$ above
is analogous to the parameter $\lambda$ labeling the pure gauge solutions 
studied by Schnabl\cite{Schnabl}. For $\lambda<1$ the solutions are 
pure gauge; for $\lambda>1$ they are ill-defined; and for $\lambda=1$ they 
describe the closed string vacuum. The parameter $\gamma$ in condition 
{\bf (iii)} is the exact counterpart of $\gamma$ in \eq{magnified}; 
thus, composite wedge solutions can be partitioned into families labeled by a 
representative pure wedge solution. The restriction $\gamma>0$ follows from
{\bf (i)} in the pure wedge case, though we are not sure if it is 
necessary in general. We assume it is true for good measure.

Let us express the conditions {\bf (ii)},{\bf (iii)} in a few other convenient
forms. Defining the square of $f(t)$,
\begin{equation} w(t) \equiv f\star f(t)\end{equation}
conditions {\bf (ii)}, {\bf (iii)} imply,
\begin{equation}\int dt w(t) = \lambda = 1\ \ \ 
\int dt\, t w(t) = \gamma \end{equation}
It is also useful to introduce the Fourier transform of $w(t)$:
\begin{equation}w(t) = \int \frac{dx}{2\pi} W(x) e^{it x}\end{equation}
{\bf (ii)} and {\bf (iii)} are then re-expressed as:
\begin{equation}W(0) = 1\ \ \ W'(0) = -i\gamma
\end{equation} Note that convolution products of $w(t)$ are isomorphic to 
local products of $W(x)$.

To calculate the energy, it is necessary to regulate the solutions 
\eq{O_fsol}. Following Schnabl\cite{Schnabl}, we rewrite them in the form,
\begin{equation}\Psi = \lim_{N\to\infty}\left(\sum_{n=0}^N \psi_n'- \psi_N
\right)\label{eq:SS_sol_reg}\end{equation}
where,
\begin{equation}\psi_n' = FcBKF^{2n}cF\ \ \ \ \ 
\psi_n = \frac{1}{\gamma}FcBF^{2n}cF\end{equation}
This is a truncated Taylor expansion of \eq{O_fsol}, modulo the $\psi_N$ 
piece which vanishes when contracted with Fock space 
states\cite{Schnabl,Okawa,Fuchs}. The factor of $1/\gamma$ in front of 
of $\psi_n$ is important. Though it is absent for Schnabl's 
solution ($\gamma=1$) it is necessary to get the right energy from 
pure wedge solutions, and composite wedge solutions in general. Note that 
generally $\psi_n' \neq \frac{d}{d n}\psi_n$, though this is the case for
pure wedge solutions.

Let us explain briefly why the $\psi_N$ term in \eq{SS_sol_reg} vanishes in 
the Fock space when $N\to\infty$. First, note that the infinite
power of any $F$ satisfying {\bf (i)}-{\bf (iii)} gives the sliver 
state:
\begin{equation}\lim_{N\to\infty} F^N = \Omega^\infty \end{equation} This can 
be readily established using a trick which we explain in the next section. 
Next, consider the limit 
\begin{equation}\lim_{N\to\infty} K F^N = K \Omega^\infty = 
\left.\frac{d}{d\alpha}\Omega^\alpha \right|_{\alpha=\infty} = 0 \end{equation}
This vanishes since when $\alpha\to\infty$ the wedge state approaches a 
constant independent of $\alpha$ (the sliver). In fact, an explicit 
computation in the level expansion reveals that 
$\frac{d}{d\alpha}\Omega^\alpha$ vanishes as $1/\alpha^3$. Since from the 
perspective of commutation with the Virasoro generators the field $K$ is 
interchangeable with $B$, we may also conclude,
\begin{equation}\lim_{N\to\infty} B F^N =0\end{equation}
Star multiplication with $FcF$ in both sides does not change this result. 
Therefore, $\psi_N$ vanishes in the large $N$ limit.

\section{Energy}
\label{sec:energy}

Now let us calculate the energy. Assuming the equations of motion, the action 
evaluated on \eq{SS_sol_reg} takes the form,
\begin{equation}S(\Psi) = -\frac{1}{6}\langle \Psi,Q_B\Psi\rangle
=-\frac{1}{6}\left(\mathcal{K}_0-2\mathcal{K}_1+\mathcal{K}_2\right)
\end{equation} 
where,
\begin{eqnarray}\mathcal{K}_0 &=& \lim_{N\to\infty}\langle \psi_N,Q_B\psi_N
\rangle\nonumber\\
\mathcal{K}_1 &=& \lim_{N\to\infty} \sum_{m=0}^N\langle \psi_N,Q_B\psi_m'
\rangle\nonumber\\
\mathcal{K}_2 &=& \lim_{N\to\infty} \sum_{m,n=0}^N\langle \psi_m',Q_B\psi_n'
\rangle\nonumber\\
&=& \lim_{N\to\infty} \sum_{m=1}^N\sum_{n=N-m+1}^N
\langle \psi_m',Q_B\psi_n'\rangle\label{eq:Ks}
\end{eqnarray}
In the second line of the third equation we have used,
\begin{equation}\sum_{k=0}^n\langle\psi'_{n-k},Q_B\psi_k'\rangle =0 
\label{eq:diag_sum}\end{equation} to cancel off the ``lower half triangle''
of the double sum. In (I) this equation was shown to hold for arbitrary choice
of $F$. To prove Sen's conjecture\cite{Sen}, we must demonstrate
\begin{equation}E = -S(\Psi) = -\frac{1}{2\pi^2}\end{equation}
in the appropriate units. 

The important consequence of \eq{diag_sum} is that we only need to calculate
BRST inner products of the $\psi_n$s when $m+n$ is large---of order $N$. Let 
us consider:
\begin{eqnarray}
\langle \psi_m,Q_B\psi_n\rangle &=& \frac{1}{\gamma^2}\int
d^3 t d^3 s \left[f(t_1)f^{\star 2m}(t_2)f(t_3)\right]
\left[f(s_1)f^{\star 2n}(s_2)f(s_3)\right]\nonumber\\
&\ &\ \ \ \ \ \ \ \ \ \ \ \ \ \ \ \ \ \ \ \ \ \ \ \ \ 
\times\Tr\left[\Omega^{t_1} cB\Omega^{t_2}c\Omega^{t_3}
d(\Omega^{s_1}cB\Omega^{s_2} c\Omega^{s_3})\right]
\end{eqnarray}
Here we use ``split string'' notation $\Tr[AB] = \langle A,B\rangle$; 
$f^{\star m}$ is the $m$th power of $f$ calculated with the convolution
product. A little manipulation brings this into the form,
\begin{equation}
\langle \psi_m,Q_B\psi_n\rangle = \frac{1}{\gamma^2}\int
dsdt\,dSdT\, [w(s)w(S)^{\star m}] [w(t)w(T)^{\star n}]C(s,S,t,T)
\label{eq:psiQpsi}\end{equation}
where,
\begin{equation}C(s,S,t,T) = \Tr\left[\Omega^s cB\Omega^Sc\Omega^t
d(cB\Omega^T c)\right]\label{eq:C}\end{equation}
This quantity can be calculated as a correlation function on the cylinder;
the full expression is complicated, but fortunately we will not need it. 

Let us now consider the limit $m+n\to\infty$ of \eq{psiQpsi}. This limit
of course is related to the limit $N\to\infty$ in \eq{Ks}---that is, $m$ and 
$n$ are typically of order $N$. We make this explicit by defining,
\begin{equation}m = aN\ \ \ \ n = bN\label{eq:sub1}\end{equation} The other ingredient we need
is a substitution of variables in the integrand \eq{psiQpsi}:
\begin{equation}S = N\s\ \ \ \ T = N\t \label{eq:sub2}\end{equation}
We then want to calculate,
\begin{equation}
\lim_{N\to\infty}\langle \psi_m,Q_B\psi_n\rangle = \frac{1}{\gamma^2}
\lim_{N\to\infty} \int
(dsdt)\,(d\s d\t) \, w(s)w(t)[N w(N\s)^{\star aN}][N w(N\t)^{\star bN}]
C(s,N\s,t,N\t)\label{eq:psiQpsi2}\end{equation}
To make sense of this, we need to understand the limit,
\begin{equation} \lim_{N\to\infty}N w(N\s)^{\star aN}\end{equation}
This expression is more transparent if we Fourier transform to ``position 
space'' $w(t)\to W(x)$:
\begin{equation} N w(N\s)^{\star aN} = \int \frac{dx}{2\pi}
W\left(\frac{x}{N}\right)^{aN} e^{i\s x}\end{equation}
Thus the limit becomes,
\begin{equation}\lim_{N\to\infty}W\left(\frac{x}{N}\right)^{aN}\end{equation}
Now note that, as a result of {\bf (ii)} and {\bf (iii)}, $W$ has a Taylor
expansion,
\begin{equation}W(x)=1 - i\gamma x + ...
\end{equation}
Plugging this in, the limit converges to an exponential:
\begin{equation}\lim_{N\to\infty}W\left(\frac{x}{N}\right)^{aN} = 
e^{-i\gamma a x}\end{equation}
or,
\begin{equation} \lim_{N\to\infty}N w(N\s)^{\star aN} = \delta(\s - a\gamma)
\end{equation} Thus, as far as this factor is concerned, in the large $N$ 
limit the function $w$ is indistinguishable from a pure wedge solution!

\Eq{psiQpsi2} now simplifies to,
\begin{equation}
\lim_{N\to\infty}\langle \psi_m,Q_B\psi_n\rangle = \frac{1}{\gamma^2}
\lim_{N\to\infty} \int
ds dt \, w(s)w(t)\, C(s,N\gamma a,t,N\gamma b)\end{equation}
To proceed we need some information about the function $C$. In appendix 
\ref{ap:correlator} we evaluate $C$ in the large $N$ limit by calculating
the relevant correlator. The result is,
\begin{equation}\lim_{N\to\infty}C(s,N\gamma a,t,N\gamma b) = st 
\mathcal{F}_K(a,b)\label{eq:C_inf}\end{equation}
where $\mathcal{F}_K$ is the function introduced in ref.\cite{Okawa},
\begin{equation}\mathcal{F}_K(a,b) = \frac{1}{\pi^2}\left(1+
\cos\frac{\pi(a-b)}{a+b}\right)+\frac{2ab}{(a+b)^2}\cos\frac{\pi(a-b)}{a+b}
+\frac{1}{\pi}\frac{a-b}{a+b}\sin\frac{\pi(a-b)}{a+b}\label{eq:F_K}
\end{equation} It is crucial that the dependence on $s$ and $t$ appears only 
in the factor $st$ in the large $N$ limit. In particular, we can calculate
\begin{equation}
\lim_{N\to\infty}\langle \psi_m,Q_B\psi_n\rangle = \frac{1}{\gamma^2}
\mathcal{F}_K(a,b)\int ds dt \, st w(s)w(t) =
\frac{1}{\gamma^2}
\mathcal{F}_K(a,b)\gamma^2 \end{equation}
or,
\begin{equation}\lim_{N\to\infty}\langle \psi_m,Q_B\psi_n\rangle = 
\mathcal{F}_K(a,b)\end{equation} This is exactly the result calculated for 
Schnabl's solution. If anything other than the factor $st$ had
appeared in $\lim_{N\to \infty}C$, the factors of $\gamma$ would not have
canceled and the solutions would violate Sen's conjecture. Note that pure
wedge solutions would have been consistent with a more general quadratic
dependence of $s,t$, so this is a nontrivial check on the consistency of 
these solutions. 

Now consider the inner product,
\begin{equation}
\langle \psi_m,Q_B\psi_n'\rangle = \frac{1}{\gamma}\int
dsdt\,dSdT\, [w(s)w(S)^{\star m}] [w(t)w(T)^{\star n}]\frac{d}{dT}C(s,S,t,T)
\label{eq:psiQpsip}\end{equation}
Making the substitutions eqs.(\ref{eq:sub1},\ref{eq:sub2}) we may compute
the large $N$ limit:
\begin{eqnarray}
\lim_{N\to\infty}\langle \psi_m,Q_B\psi_n'\rangle &=& \frac{1}{\gamma}
\lim_{N\to\infty} \frac{1}{N}\int
(dsdt)(d\s d\t)\, w(s)w(t)[N w(N\s)^{\star aN}] [Nw(N\t)^{\star bN}]
\frac{\d}{\d\t}C(s,N\s,t,N\t)\nonumber\\
&=& \frac{1}{\gamma}
\lim_{N\to\infty} \frac{1}{N}\int
(dsdt)(d\s d\t)\, w(s)w(t)\delta(\s-a\gamma)\delta(\t-b\gamma)
\frac{\d}{\d\t}C(s,N\s,t,N\t)\nonumber\\
&=& \frac{1}{\gamma}
\lim_{N\to\infty} \frac{1}{N} \frac{\d}{\d(\gamma b)}\mathcal{F}_K(a,b)
\int dsdt\, st w(s)w(t)\nonumber\\
&=&\lim_{N\to\infty} \frac{1}{N}\frac{\d}{\d b}\mathcal{F}_K(a,b)
\end{eqnarray} Again, the factors of $\gamma$ cancel out and we get the same
result as for Schnabl's solution. A similar argument shows,
\begin{equation}
\lim_{N\to\infty}
\langle \psi_m',Q_B\psi_n'\rangle = \lim_{N\to\infty}\frac{1}{N^2}
\frac{\d}{\d a}\frac{\d}{\d b}\mathcal{F}_K(a,b)\end{equation}
Calculation of the energy now proceeds exactly as it does for Schnabl's 
solution\cite{Schnabl,Okawa,Fuchs}. The factors of $1/N,1/N^2$ turn the
sums over $\psi_n'$ in $\mathcal{K}_1,\mathcal{K}_2$ into Riemann integrals
over $a,b$. Thus,
\begin{eqnarray}E &=& \frac{1}{6}(\mathcal{K}_0-2\mathcal{K}_1+\mathcal{K}_2)
\nonumber\\
&=& \frac{1}{6}\left(\mathcal{F}_K(1,1)
+2\int_0^1 db\frac{d}{db}\mathcal{F}_K(1,b)+\int_0^1 da\int_{1-y}^1 db
\frac{\d}{\d a}\frac{\d}{\d b}\mathcal{F}_K(a,b)\right)\nonumber\\
&=& \frac{1}{6}\left[\left(\frac{2}{\pi^2}+\frac{1}{2}\right) - 
2\left(\frac{2}{\pi^2}+\frac{1}{2}\right) +\left(-\frac{1}{\pi^2}+\frac{1}{2}
\right)\right]\nonumber\\
&=& -\frac{1}{2\pi^2}\end{eqnarray}
Therefore, provided $f(t)$ satisfies conditions {\bf (i)}, {\bf(ii)}, and 
{\bf (iii)}, composite wedge solutions all describe the endpoint of tachyon
condensation, i.e. the closed string vacuum. Following the discussion of 
refs.\cite{Okawa,Fuchs} we can also show that the equations of motion are still
satisfied when contracted with the solutions themselves. 

\section{Cohomology}
\label{sec:cohomology}

An acceptable solution for vacuum must satisfy one other important constraint:
its linearized fluctuations should support no physical open string 
excitations. For composite wedge solutions, the proof of this is simple and 
worth demonstrating. We must argue that the shifted BRST operator,
\begin{equation}d_\Psi = d + [\Psi,\cdot]\end{equation}
has vanishing cohomology. Following refs.\cite{cohomology1,cohomology2}, this
follows if we can construct a field $A$ (the ``homotopy operator'') such that
\begin{equation}d_\Psi A = 1 \end{equation}
We now suppose that $A$ can be found within the subalgebra generated by $K,B$ 
and $c$. Since $A$ has ghost number $-1$, its form is fixed to be: 
\begin{equation}A = BG(K),\end{equation}
where $G$ is some field which depends only on $K$. For the moment it is 
sufficient to consider the unregularized solution, \eq{SSsol}; we compute,
\begin{equation}d_\Psi A = KG + F cB \frac{F}{1-F^2}KG + KG\frac{F}{1-F^2}BcF
\end{equation}
If we want the $B$s and $c$s to cancel, we should {\it define} $G$ so that it 
satisfies,
\begin{equation} KG = 1-F^2 \label{eq:G_def}\end{equation}
Plugging this in,
\begin{equation}d_\Psi A = 1-F^2 + F (cB+Bc) F = 1
\end{equation}
proving that $d_\Psi$ has no cohomology.

This proof assumes that \eq{G_def} has a solution for $G$ which yields a 
well-defined homotopy operator. To check this, let us solve \eq{G_def} 
explicitly. Writing $G$ in the form,
\begin{equation}G = \int_{-\infty}^\infty dt g(t) \Omega^t \end{equation}
we can compute,
\begin{eqnarray}KG &=& \int_{-\infty}^\infty dt g(t) \frac{d}{dt}\Omega^t
\nonumber\\ &=& 
g(\infty)\Omega^\infty - \int_{-\infty}^\infty dt \frac{d}{dt} g(t) \Omega^t
\nonumber\\ &=& 1-F^2 \end{eqnarray}
Since $1-F^2$ has no term proportional to the sliver, we can only find 
a solution if $g(t)$ vanishes at infinity\footnote{Actually, if we want the 
field $G$ to converge, we should require that $g(t)$ is an integrable function 
on the real line; this is stronger than requiring that $g$ vanishes at 
infinity. However, really what we want is the homotopy operator $=BG$ to 
converge. This constraint is much weaker; in fact, $g(t)$ could even blow up
at infinity, as long as the divergence is slower than quadratic.}:
\begin{equation}g(\infty) = 0\end{equation}
\Eq{G_def} now becomes a differential equation for $g(t)$:  
\begin{equation}\frac{d}{dt}g(t) = w(t)-\delta(t) \label{eq:diff_goft}
\end{equation}
Integrating,
\begin{equation}g(t) = \left\{\begin{matrix} 0\ \ \ \ \ & \mathrm{for}\ t<0 \\
\displaystyle{-1 +\int_0^t ds\, w(s) } 
\ \ \ & \mathrm{for}\ t>0\end{matrix}\right.\label{eq:goft}
\end{equation}
Consistency demands that this expression vanish at infinity,
\begin{equation}g(\infty) = -1+\int_0^\infty ds\, w(s) = -1+\lambda = 0
\end{equation} This is precisely condition {\bf (ii)} needed to get 
nontrivial energy out of our solutions. Therefore, the 
homotopy operator $A$ exists, and the physical spectrum is empty, only for 
those solutions whose energies match that of a decayed D-brane. For Schnabl's 
solution $w(t) = \delta(t-1)$ the homotopy operator takes the form,
\begin{equation}A = -B\int_0^1 dt \Omega^t\end{equation}
in agreement with the result of Ellwood and Schnabl\cite{cohomology2}.

For simplicity, in the above discussion we have used the unregularized solution
and ignored possible terms, like the $\psi_N$ piece, which vanish in the 
Fock space. We are uncertain how such terms would effect the cohomology
if present, but at any rate a quick calculation with the regularized solution
\eq{SS_sol_reg} and our expression for $G$ \eq{goft} reveals that 
$d_\Psi A = 1$ even outside the Fock space, as was demonstrated for the 
case of Schnabl's solution in ref.\cite{cohomology2}. Actually, this is less 
of a stringent consistency check than a resolution to an ambiguity in the 
definition of $A$ outside the Fock space; in particular, \eq{G_def} only 
determines the homotopy operator up to a term proportional to the sliver,
\begin{equation}A = B(G + u\Omega^\infty)\end{equation} 
since the sliver is annihilated by $K$ and does not contribute to \eq{G_def}.
If we require $d_\Psi A=1$ even up to such ``vanishing terms,'' the 
undetermined constant $u$ is fixed to be zero and \eq{goft} still gives the 
full solution for the homotopy operator.

\section{Conclusion}

In this paper we have constructed an infinite class of distinct and nontrivial
descriptions of the closed string vacuum. It would be interesting to use these
solutions to try to better understand the nature of open string gauge 
symmetry---specifically, to understand which gauge transformations are 
allowed and which are not allowed, and for what reasons. Recall that, at 
least naively, all the split string solutions \eq{SSsol} are gauge equivalent 
to $\Psi=0$. However, apparently only a subset of the gauge transformations
preserving conditions {\bf (i)}, {\bf (ii)}, and {\bf (iii)} are actually 
valid.

We have not attempted a computation of the energies of these 
solutions in level truncation, though such calculations would be interesting. 
It seems likely that the convergence of the energy will vary widely depending 
on the choice of $f(t)$. For example, consider
\begin{eqnarray}f(t) &=& \delta(t) - \frac{1}{2}\delta'(t)\nonumber\\
F &=& 1+\frac{1}{2}K\label{eq:id_sol}
\end{eqnarray} This is an identity based solution whose energy is probably not
very convergent in level truncation. Still it satisfies  {\bf (i)} 
(marginally), {\bf (ii)}, and {\bf (iii)}---in fact, from the perspective of 
our analytic calculation, \eq{id_sol} is in the same class as Schnabl's 
solution with $\gamma = 1$. It is also worth pointing out that condition 
{\bf (i)}, while it seems necessary, does not appear to play a role in the 
evaluation of the energy. Probably the hitch here is \eq{diag_sum}, which 
allowed us to cancel off the $\psi_n$ inner products for small $m+n$. For 
solutions which violate or only marginally satisfy {\bf (i)}, the $\psi_n$ 
inner products for small $m+n$ are not very well defined, so \eq{diag_sum} 
should be treated with some suspicion. The other question is the necessity of
condition {\bf (iii)}. If the other conditions are satisfied, it seems the 
energy calculation goes unchanged---$\gamma$ cancels out even if it is 
negative. Our only concern is that for $\gamma<0$, the function $C$ is 
formally an inner product of inverse wedge states. However, the analytic 
continuation to $\gamma<0$ appears fine, so these solutions could be 
okay after all.

To us, it is highly surprising that Schnabl's regularization \eq{SS_sol_reg}
remains valid for composite wedge solutions. In some sense the regularization
is ``unnatural,'' since each $\psi_n$ is composed of wedge states of all
sizes. It would be interesting to understand this regulator better, or to 
discover other ways of regulating the solutions and calculating their 
energies.

The author would like to thank A. Sen for thoughtful conversations. This
work was supported by the Department of Atomic Energy, Government of India.

\begin{appendix}
\section{Correlator}
\label{ap:correlator}

\begin{figure}
\begin{center}
\resizebox{2.6in}{2in}{\includegraphics{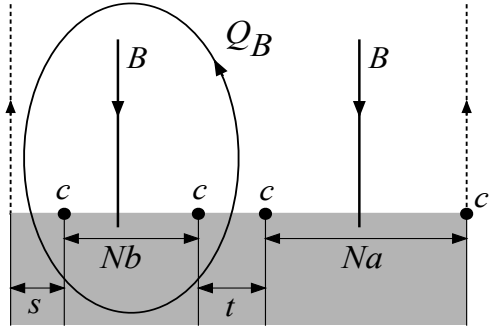}}
\end{center}
\caption{\label{fig:C} Correlation function defining $C(s,Na,t,Nb)$.}
\end{figure}

In this appendix we would like to evaluate \eq{C},
\begin{equation}C(s,aN,t,bN) = \Tr\left[\Omega^s cB\Omega^{aN}c\Omega^t
d(cB\Omega^{bN} c)\right]\end{equation}
in the large $N$ limit. This quantity can be calculated as a correlation 
function on the cylinder, as shown in figure \ref{fig:C}
\begin{equation}C(s,aN,t,bN)= \left\langle c(t+s+N(b+a))Bc(t+s+Nb)
\left(\oint\frac{dz}{2\pi i} 
j_B(z)c(s+Nb)Bc(s)\right)\right\rangle_{C_{t+s+N(b+a)}}\end{equation}
The BRST current encircles all operators in the large parentheses, and 
(as usual) the $B$ contours pass between the $c$ ghosts on either side.
Calculating the BRST variation and moving around some $B$ contours, this
can be reduced to a sum of five terms:
\begin{equation}C = X_1+X_2+X_3+X_4+X_5\end{equation}
where,
\begin{eqnarray}X_1 &=& 
\langle c(t+s +N(a+b))c(t+s+Nb)c\d c(s+Nb)B\rangle_{C_{t+s+N(b+a)}}\nonumber\\
X_2 &=& 
-\langle c(t+s+Nb)c\d c(s+Nb)B c(s)\rangle_{C_{t+s+N(b+a)}}\nonumber\\
X_3 &=& 
-\langle c(t+s+N(b+a))Bc(t+s+Nb)c(s+Nb)K 
c(s)\rangle_{C_{t+s+N(b+a)}}\nonumber\\
X_4 &=& 
\langle c(t+s+N(b+a))Bc(s+Nb)c\d c(s)\rangle_{C_{t+s+N(b+a)}}\nonumber\\
X_5 &=& 
-\langle c(t+s+N(b+a))Bc(t+s+Nb)c\d c(s)\rangle_{C_{t+s+N(b+a)}} \end{eqnarray}

Let us now concentrate on $X_1$. To understand the $N\to\infty$ limit it is 
helpful to scale the correlator by $1/N$. Scaling produces a factor of
$N$ for each $c$, no factor for $\d c$, and a factor of $1/N$ for the $B$ 
contour. All in all, $X_1$ becomes,
\begin{equation}X_1 = N^2\left\langle c\left(a+b+\frac{s+t}{N}\right)
c\left(b + \frac{s+t}{N}\right)c\d c\left(b+\frac{s}{N}\right)
B\right\rangle_{C_{a+b+\frac{s+t}{N}}}\end{equation}
There is an $N^2$ divergence in front, but note that a $c$ and $c\d c$
become coincident for large $N$, which contributes a corresponding vanishing 
factor. Specifically,
\begin{eqnarray}c\left(b + \frac{s+t}{N}\right)c\d c\left(b+\frac{s}{N}\right)
&=&\left[c + \frac{t}{N}\d c +
\frac{1}{2}\left(\frac{t}{N}\right)^2\d^2 c+...\right]c\d c\left(b+\frac{s}{N}
\right)\nonumber\\
&=&\frac{1}{2}\left(\frac{t}{N}\right)^2 c\d c\d^2c\left(b+\frac{s}{N}\right)+
\mathcal{O}(N^{-3})\end{eqnarray} 
Plugging this in,
\begin{equation}\lim_{N\to\infty}X_1 = \frac{1}{2}t^2\left\langle c(a+b)c\d 
c\d^2c(b)B\right\rangle_{C_{a+b}}\end{equation}
A similar argument for $X_2$ shows,
\begin{equation}\lim_{N\to\infty}X_2 = -\frac{1}{2}t^2\left\langle c\d 
c\d^2c(b)B c(0)\right\rangle_{C_{a+b}} = -\lim_{N\to\infty}X_1\end{equation}
In particular, $X_1$ and $X_2$ cancel in the large $N$ limit. Indeed this
is fortunate, since as discussed in the text a $t^2$ dependence in the 
large $N$ limit (or any dependence other than $st$) would imply composite
wedge solutions violate Sen's conjecture. A similar argument shows that 
$X_4$ and $X_5$ cancel.

This leaves the term $X_3$, which in the infinite $N$ limit is:
\begin{equation}\lim_{N\to\infty}X_3 = 
-st\langle c\d c(a+b)B c\d c(b) K\rangle_{C_{a+b}}\label{eq:X3_inf}
\end{equation} The factor of $s$ comes from the OPE of the two $c$s near 
$a+b$ and the factor of $t$ comes from the OPE near $b$. The above 
correlator now basically has to be $-\mathcal{F}_K(a,b)$ from \eq{F_K}. 
For completeness let us explain how this comes about. Recalling that the
$K$ insertion can be calculated as a derivative, a little rearrangement 
brings \eq{X3_inf} to the form,
\begin{equation}\lim_{N\to\infty}X_3 = 
-st\left.\frac{\d}{\d b}\frac{\d}{\d x}\frac{\d}{\d y}
\langle c(a+b)c(x)c(a)c(y) B\rangle_{C_{a+b}}\right|_{x=a+b,y=a}
\label{eq:X3_inf2}\end{equation}
This is a well-known correlator\cite{Schnabl,Okawa}:
\begin{eqnarray}\langle c(x_4)c(x_3)c(x_2)c(x_1)B\rangle_{C_L}
&=& \frac{L^2}{\pi^3}\left[
x_1\sin\frac{\pi x_{43}}{L} \sin\frac{\pi x_{42}}{L} \sin\frac{\pi x_{32}}{L} 
-x_2\sin\frac{\pi x_{43}}{L} \sin\frac{\pi x_{41}}{L} \sin\frac{\pi x_{31}}{L} 
\right.\nonumber\\ &\ &\ \ \ \ \ \left.
+x_3\sin\frac{\pi x_{42}}{L} \sin\frac{\pi x_{41}}{L} \sin\frac{\pi x_{21}}{L} 
-x_4\sin\frac{\pi x_{32}}{L} \sin\frac{\pi x_{31}}{L} \sin\frac{\pi x_{21}}{L} 
\right]\nonumber\\
\end{eqnarray}
where $x_{ij}=x_i-x_j$. We actually found it easier to work with another
form,
\begin{eqnarray}\langle c(x_4)c(x_3)c(x_2)c(x_1)B\rangle_{C_L}
&=& \frac{L^2}{4\pi^3}\left[x_{12}\sin\frac{2\pi x_{43}}{L} 
+ x_{31}\sin\frac{2\pi x_{24}}{L} + x_{43}\sin\frac{2\pi x_{12}}{L} 
\right.\nonumber\\ &\ & \ \ \ \ \ \left.
+ x_{24}\sin\frac{2\pi x_{13}}{L} + x_{32}\sin\frac{2\pi x_{14}}{L} 
+ x_{14}\sin\frac{2\pi x_{32}}{L}\right]\nonumber\\\end{eqnarray}
which is manifestly translationally invariant and linear in sines. Plugging 
this in to \eq{X3_inf2} quickly yields,
\begin{equation}\lim_{N\to\infty}C(s,aN,t,bN) = st\mathcal{F}_K(a,b)
\end{equation}
consistent with \eq{C_inf}.

\end{appendix}

\end{document}